\documentclass[a4paper,10pt]{article}

\usepackage{latexsym}
\usepackage{graphicx}
\usepackage{amsmath}

\title{Non-Steady wall-bounded flows of viscoelastic fluids under periodic forcing.}
\author{Anier Hern\'andez-Garc\'ia$ ^1 $, Antonio Fern\'andez Barbero$ ^2$, and \and Oscar Sotolongo-Costa$ ^1$}

\begin{document}
\date{}
\maketitle
\begin{center}
 \itshape{$\;\ ^1$"Henri Poincar\'e" Group of Complex Systems,  University of Havana,  Cuba
\\$\;\ ^2$Group of Complex Fluids, Almeria University, Spain
}
\end{center}


\begin{abstract}
 The problem of oscillating flows inside pipes under periodic forcing of viscoelastic fluids is addressed here. Starting from the linear Oldroyd-B model, a generalized Darcy's law is obtained in frequency domain and an explicit expression for the dependence of the dynamic permeability on fluid parameters and forcing frequency is derived. Previous results in both viscoelastic and Newtonian fluids are here shown to be particular cases of our results. On the basis of our calculations, a possible explanation for the observed damping of local dynamic response as the forcing frequency increases is given. Good fitting with recent experimental studies of wave propagation in viscoelastic media is here exhibited. Sound wave propagation in viscoelastic media flowing inside straight pipes is investigated. In particular, we obtain the local dynamic response  for weakly compressible flows. 
\end{abstract}

\section{Introduction}
  Wall-bounded oscillating flows of Newtonian fluids driven by periodic forcing have been extensively studied both theoretically and experimentally. Interesting phenomena such us the formation of non-steady boundary layers have been observed, provided that the frequency of the applied force is larger than the inverse of the characteristic time in which the vorticity diffuses along the cross-section (see for instance \cite{Schlichting} and references therein). Recently, the formation of a non-steady boundary layer has also been predicted for Oldroyd-B fluids \cite{rcf}.

 In recent advances in many topics ranging from soft-matter, biofluid mechanics, rocket propulsion, rheology among others are encountered flows in which Newtonian  approximation breaks down. Remarkably, it has been recently observed in oscillating wall-bounded flows of these fluids that when relaxation times of elastic stresses are comparable with characteristics  times scales of the flow, the nontrivial interactions of elastic degrees of freedom with the inertial ones and flow geometry give rise to a resonant like behavior that is absent in the corresponding Newtonian flow (\cite{enhacement},\cite{experiments},\cite{pre2}). Thus, the description of oscillating wall-bounded flows of non-Newtonian fluids is of great importance for both fundamental physics and applied topics.  

  In Refs. (\cite{experiments},\cite{pre2}), by means of velocity measurements at the center of a viscoelastic fluid column ($NayCl/NaSal$ \textit{in water}) flowing inside a straight pipe of circular cross section \cite{experiments} and more recently  in the entire cross section of the fluid column \cite{pre2}, it was found a dramatic enhancement in the dynamic response to an oscillating periodic pressure gradient, that is, the fluid response, measured in terms of the velocity for a given amplitude of the pressure gradient, exceeds by several orders in magnitude that obtained for the steady case. It is noteworthy that most of the features observed in experiments, namely, the values of the driving frequencies at which the resonant behavior takes place and the formation of Couette-like flows  as the driving frequency increases are successfully reproduced by a linear Navier-Stokes and Maxwell constitutive equations (\cite{enhacement},\cite{pre2},\cite{pre1}). Nevertheless, theoretical predictions within this model overestimate the magnitude of the response amplitude as well as the root mean square velocity at the pipe axis and the velocity at certain points of the cross section of the fluid column. Furthermore, the Maxwell model cannot describe the damping observed in the peaks of dynamic permeability as the driving frequency is increased (at least in the range of frequencies covered by both experiments). These issues motivated us to extend earlier theoretical works in the hope that this may shed light on the the physical mechanisms that could lead to the observed phenomena.    
         
 In this article, specifically, we shall try to assess the effects that pure Newtonian contributions (due to the solvent) to the total stress tensor of viscoelastic solutions will have on the fluid response in wall-bounded oscillating flows under periodic forcing. In order to achieve that, we shall assume that stresses in the viscoelastic solution obey the  linear Oldroyd-B constitutive equation (\cite{Oldroyd}, \cite{Skelland},\cite{birdstewart}).  In particular, we present a detailed theoretical derivation of the velocity field and the dynamic permeability in flows at small both Reynolds and Weissenberg numbers. As we shall see, our finding suggests that the (additional) Newtonian component of stresses due to solvent viscosity might leads to the damping of the local dynamic response as the forcing frequency increases, a fact observed in (\cite{experiments},\cite{pre2}). We also investigate the behavior of these quantities in weakly compressible flows. Our main purpose is compare our findings against known results for incompressible fluids.
 The article is organized as follows. In section II, for completeness, we briefly review main aspects of the Oldroyd-B model. In section III we introduce the main assumptions and the resulting set of governing equations we have used to derive explicit expressions for the dynamic permeability, the velocity field and the local dynamic response which are defined  within this section. Here we also compare our results with the measured values of the local dynamic response and the root mean square velocity reported in \cite{experiments}. In section IV, the problem of modeling stresses for weakly compressible flows is revisited and starting from it, a explicit expression for the local dynamic response is obtained. A comparison of the latter magnitude against the corresponding incompressible case, bearing in mind the experimental conditions, is also made.

 Some of the definitions and expressions presented in this article have already been presented in previous publications. Wherever needed, we shall repeat them for the sake of clarity and completeness.

\section{Oldroyd-B model}
 One of the simplest models that  describes basic viscoelastic behavior of an incompressible liquid having a constant shear viscosity, and constant relaxation and retardation  times is the so-called Oldroyd-B model (\cite{Oldroyd}, \cite{Skelland},\cite{birdstewart}). This is an extension of the Maxwell model that fits quite well data from polymeric solutions.

  In the Oldroyd-B model, it is considered that viscoelastic fluids can be regarded as a dilute suspension of elastic dumbbells. Thus, the complicated structure of the molecules that form the liquid is replaced by a simpler mechanical structure consisting of identical pairs of microscopic beads (of negligible mass) connected by Hookean springs. Besides, the concentration is supposed to be uniform and very low. The latter ensures that polymer-polymer interactions are negligible. Furthermore, the viscoelastic solution is regarded as a continuous medium. The "coarsening" involved in such description requires to get rid of the microscopic degrees of freedom. Consequently, the reaction of the dumbbells on the fluid is treated at a mean field level described by an elastic contribution $\sigma^e$ to the total stress tensor, which is found to be proportional to the conformation tensor, i.e. 

\begin{equation}
 \sigma^e_{ij}\propto <R_i R_j> ,
\label{polymercontribution}  
\end{equation}
where $R_i$ denotes the i-th component of the elongation vector (obtained by subtracting the position of one bead from that of the other one). In equation \ref{polymercontribution} the average is taken over the statistics of the thermal noise, or equivalently over a volume large enough to contain a huge number of molecules but small compared with the whole fluid volume. The conformation tensor is defined as $\sigma^c_{ij}=\frac{<R_iR_j>}{R_0^2}$,  where $R_0$ is the equilibrium length of the spring. 
  The evolution equation for the conformation tensor can be inferred by considering the forces which act on the beads, namely, the hydrodynamic drag which obeys a Stokes' law, a Brownian force and the elastic spring force. When details of the kinetic theory are worked out one gets the following model incorporating the elastic nature of the dumbbells stress tensor (\cite{birdstewart},\cite{larsonins},\cite{Prilutski}), the so-called Maxwell model

\begin{equation}
 \sigma^e+\lambda\frac{\delta \sigma^e}{\delta t}=\eta_p\left[\nabla v+(\nabla v)^t\right],
\label{maxwellinv}  
\end{equation}
where $\eta_p$ is the contribution of the elastic additives to the total shear viscosity in steady flows at small shear rates, $(\nabla v)_{ij}=\frac{\partial v_i}{\partial x_j}$ is the velocity gradient tensor, $(\nabla v)^t$ is its transpose and $\lambda$ is the relaxation time of elastic stresses. The symbol $\frac{\delta }{\delta t}$ denotes the upper-convected time derivative defined by (\cite{birdstewart},\cite{larson})

\begin{equation}
 \frac{\delta G}{\delta t}=\frac{\partial G}{\partial t}+(v\cdot\nabla)G-\left[G\cdot\nabla v+ (\nabla v)^t\cdot G\right]
\label{upperconvected}
\end{equation}
 
 Thus, the above time derivative takes into account that the relation between stresses and kinematic tensors at a fluid particle are independent of the instantaneous orientation of that particle. One can see that the ratio between the terms inside brackets in the r.h.s of equation \ref{upperconvected} to the linear relaxation term is determined by the dimensionless expression $\lambda \frac{V}{L}$, called the Weissenberg number, where $V$ and $L$ are a typical velocity and length scale of the flow, respectively. For instance, in a flow inside a straight pipe those magnitudes could be taken as the maximum value of the velocity across the cross section and the radius, respectively. We may also note that the ratio between the advection term $(v\cdot\nabla)G$ and the relaxation one is given by the Weissenberg number as well. Thus, when the latter number is sufficiently small the relaxation of elastic stresses overcomes the effects caused by nonlinearities in the constitutive equation [Eqs. \ref{maxwellinv} and \ref{upperconvected}]. In this limit, the upper-convected derivative could be replaced by the local time derivative
 
\begin{equation}
 \frac{\delta G}{\delta t}\simeq \frac{\partial }{\partial t}.
\label{approach}
\end{equation}
   
 In the Oldroyd-B model the total stress tensor $\sigma$ is given by the sum of a Newtonian solvent contribution $\sigma^s$ and the elastic additives contribution $\sigma^e$, i.e.

\begin{equation}
 \sigma=\sigma^e+\sigma^s.
\end{equation}

   The constitutive equation for the solvent is given by
\begin{equation}
 \sigma^s=\eta_s\left[\nabla v+(\nabla v)^t\right], 
\end{equation}
where $\eta_s$ is the solvent viscosity.

When the two contributions are added, the result is found to be
             \begin{equation}
              \sigma +\lambda\dfrac{\delta \sigma}{\delta t}=
\eta_0\left(\mathbf{D}+\lambda\frac{\eta_{s}}{\eta_0}\dfrac{\delta
\mathbf{D}}{\delta t}\right),
                \label{ecconstupper}
             \end{equation}
where $\mathbf{D}=\nabla v+(\nabla v)^t$ is the symmetric part of the velocity gradient tensor, and $\eta_0= \eta_{s}+\eta_{p}$. 

 From the above expression the Maxwell's model is recovered making $\eta_{s}\rightarrow 0$. Moreover, if we let $\lambda\rightarrow 0$, the Newtonian model is obtained.

\section{Non steady flow of Oldroyd-B liquids in straight cylindrical pipes.}
\subsection{Set of governing equations}

 We shall now proceed to derive an analytical expression of the velocity field of an Oldroyd-B liquid  subject to an oscillatory pressure gradient  and  confined in a pipe of
 uniform circular cross section.

In order to achieve it  we must solve the Navier-Stokes equation for incompressible 
fluids, i.e.
 \begin{equation}
  \rho \left(\dfrac{\partial \vec{v}}{\partial t} + (\vec{v}\cdot\nabla) \vec{v} \right)=-\nabla p +\nabla\cdot\mathbf{\sigma} 
 \end{equation}
\begin{equation}
 \nabla\cdot \vec{v}=0
\end{equation}
 with $\sigma$ given by the constitutive equation \ref{ecconstmod}. This is a highly nonlinear set of coupled equations. Hence, it turns out to be rather difficult to carry out an analytical treatment of any flow property of a viscoelastic fluid.   

  Henceforth, we shall consider only flows in which the Weissenberg and Reynolds numbers are very small. The latter means that the convective term $(\vec{v}\cdot\nabla) \vec{v} $ in the Navier-Stokes equation is negligible. With these assumptions the set of governing equations is determined by

 \begin{equation}
  \rho \dfrac{\partial \vec{v}}{\partial t}=-\nabla p +\nabla\cdot\mathbf{\sigma}, 
 \end{equation}
\begin{equation}
 \nabla\cdot \vec{v}=0
\end{equation}
and
\begin{equation}
 \sigma +\lambda\dfrac{\partial \sigma}{\partial t}=2
\eta_0\left(\mathbf{D}+\lambda\frac{\eta_{s}}{\eta_0}\dfrac{\partial
\mathbf{D}}{\partial t}\right).
\label{ecconstmod}
\end{equation}

 A more precise discussion of the above assumptions comes in the following. 
 This type of flow occurs, typically, under  the influence of a reciprocating piston located at one end of a pipe of circular cross-section (\cite{Schlichting},\cite{experiments}, \cite{pre2} and references therein). We assume that  pipe length is much larger than its radius and we shall study the flow very far from the moving piston. There, under the previous assumptions, the flow can be described by the component of the velocity along the pipe axis. Moreover, owing to the incompressibility assumed and symmetries involved, the velocity only depends on the distance to the pipe axis,  $\vec{v}= v_{z}(r,t)\mathbf{e_{z}}$ and the pressure only varies
 in the longitudinal direction $\nabla p= \frac{dp(z,t)}{dz}\mathbf{e_{z}}$. With these assumptions the  pressure gradient must be constant along 
the pipe. 

%
%
%
%
%
%

Then, the convective term in the Navier-Stokes equation  vanishes identically leading to

 \begin{equation}
 \rho \dfrac{\partial v_{z}(r,t)}{\partial t} =-\nabla{p}+\nabla\cdot\sigma.
\label{linearns}
 \end{equation}

Taking divergence in both sides of equation (\ref{ecconstmod}), from  the definition of
 $\mathbf{D}$, and bearing in mind the expression for the velocity, it can be obtained
$2\nabla\cdot\mathbf{D}=\nabla^{2}v_z$.  These assumptions lead equation \ref{ecconstmod} to
 \begin{equation}
  \nabla\cdot \sigma +\lambda\dfrac{\partial \nabla\cdot\sigma}{\partial t}=
\eta_{o}\left(\nabla^{2}v_z+\lambda_{r}\dfrac{\partial
\nabla^{2}v_z }{\partial t}\right),
                \label{divcconstmod}
 \end{equation}
where $\lambda_{r}=\lambda\frac{\eta_s}{\eta_0}$ is a constant with dimensions of time.

  If we derive with respect to time, multiply by $\lambda$  both sides of (\ref{linearns}) 
and substituting the term $\lambda\dfrac{\partial \nabla\cdot\sigma}{\partial t}$ from (\ref{divcconstmod}) we obtain
 \begin{equation}
 \lambda \rho \dfrac{\partial^{2} v_{z}(r,t)}{\partial t^{2}}+\rho \dfrac{\partial v_{z}(r,t)}{\partial t}=-\lambda\dfrac{\partial \nabla p}{\partial t}+\eta_{o}\left(\nabla^{2}v_z+\lambda_{r}\dfrac{\partial
\nabla^{2}v_z }{\partial t}\right)
\label{linearnsdesp}
 \end{equation}
%
Let us now refer all linear dimensions to the pipe radius $R$,  time to $\frac{ R^2}{\nu_{o}}$ and the  velocity to $\frac{K R^{2}}{\nu_{o}}$ where K is the amplitude of the pressure gradient divided by the fluid density and $\nu_{o}$ is its kinematic viscosity at steady motion.

Thus, it can be obtained
\begin{equation}
 A \dfrac{\partial^{2} \Lambda(\xi,\tau)}{\partial
\tau^2}+\dfrac{\partial \Lambda(\xi,\tau)}{\partial
\tau}=\nabla^{2}\Lambda(\xi,\tau)+
A\frac{\eta_{s}}{\eta_{o}}\dfrac{\partial\nabla^{2}\Lambda}{\partial
\tau}-\left(A\dfrac{\partial \nabla p}{\partial \tau} +\nabla p\right).
\label{dimensionless}
\end{equation}
where $A=\frac{\lambda \nu_{0}}{R^{2}}$ denotes the inverse of the Deborah's number,
 $\tau$, $\xi$ and  $\Lambda(\xi,\tau)$ denote the dimensionless time, radial coordinate and velocity, respectively.

Fourier transformation of the latter equation leads to

\begin{equation}
 \frac{1}{\xi}\frac{\partial}{\partial \xi}(\xi\dfrac{\partial
\widetilde{\Lambda}(\xi,\alpha)}{\partial \xi})+i \alpha \frac{1-i
\alpha A}{1-i \alpha
A\eta'}\widetilde{\Lambda}(\xi,\alpha)=\frac{1-i \alpha A}{1-i
\alpha A\eta'}\frac{\partial \widetilde{p}(\alpha)}{\partial z}
\label{difreq}
\end{equation}
where $\alpha=\frac{\omega R^{2}}{\nu_{o}}$. This number is  the ratio between  inertia and viscosity forces,  so it can be regarded as a Reynolds number for non steady flows. $\widetilde{\Lambda}(\xi,\alpha)$ and $\frac{\partial \widetilde{p}(\alpha)}{\partial z}$ 
denote the Fourier transform of the velocity and pressure gradient, respectively. Hereafter, we shall drop the upper hat and the dependence on $\alpha$ will indicate the Fourier Transform, unless otherwise noted.  In equation (\ref{difreq}) $\eta^{'}=\frac{\eta_s}{\eta_o}$  is the viscosity ratio. 

 To obtain the velocity field,  boundary conditions must be added,
 namely, the no slip at the wall and the fact that the velocity must
 remain finite at the pipe axis.

   Then it follows for the velocity

\begin{equation}
 \Lambda (\xi,\alpha)=\frac{1}{ i \alpha}\left(1-\frac{J_{0}(\beta^{'} \xi)}{J_{0}(\beta^{'} 
 )}\right) \dfrac{d p(\alpha)}{dz},
\label{vincompresible}
\end{equation}

\begin{equation}
\beta^{'2}(\alpha)=i \alpha\frac{1-i \alpha A}{1-i \alpha A \eta^{'}}
\label{argument}
\end{equation}

If we make $\eta^{'}\rightarrow 0$ in the above equation the same result as for the Maxwell model is
 obtained, (see Refs. [\cite{enhacement},\cite{experiments}, \cite{Maxwellteorico}]). In addition, 
making also $A\rightarrow 0$ the Newtonian behavior  is recovered, (see \cite{Schlichting}).

Now, following the same procedure as in Refs. (\cite{enhacement},\cite{experiments}, \cite{pre2})
we shall calculate the average velocity over a cross section. 
 Thus,

  \begin{equation}
   \left\langle \Lambda (\xi,\alpha)\right\rangle _{\xi}=\left\langle\frac{1}{i\alpha}\left(1-\frac{ 
J_{0}(\beta^{'} \xi)}{J_{0}(\beta )}\right)\right\rangle_{\xi}\dfrac{dp(\alpha)}{dz},
  \end{equation}
where the symbol $\langle(...)\rangle_{\xi}$ denotes the spatial average over the cross-section of the quantity inside brackets.

From the above expression it can be inferred that the total flux is proportional to
 the pressure gradient. This  resembles Darcy's Law in frequency domain. 
Then, we define the dynamic permeability as 

\begin{equation}
 K(\alpha)=-R^{2}\frac{\langle \Lambda (\xi,\alpha)\rangle_{\xi}}{d p(\alpha)/dz} 
\label{permeabilidad}
\end{equation}

\subsection{Comparison with experimental results}
To compare  with experimental results we shall,
 as in \cite{experiments}, define a local dynamic response (\textbf{LDR}) as follows

\begin{equation}
 \kappa(\alpha) =-R^{2}\frac{\Lambda (0,\alpha)}{dp(\alpha)/dz}
\label{localdynamic}
\end{equation}

 In Ref. \cite{experiments} a detailed experimental study of the dynamic response of 
a Newtonian fluid, the  Glycerol, and a viscoelastic fluid, \textit{CPyCl/NaSal} solution,  under an oscillating pressure gradient was
 performed. Measurements of fluid particles velocity and the root mean square velocity $V_{rms}$ were made inside a  straight vertical cylinder with circular cross section.

  In Ref. \cite{experiments} the pressure gradient was described by 
\begin{equation}
 \frac{d p(t)}{d z}=\rho z_0 w^{2} sin(\omega t) 
\label{pressuremex}
\end{equation}
in which $z_0$ represents the piston displacement amplitude equal to $0.8 mm$. Thus the experimental value of the  (\textbf{LDR}) $\kappa^{e}(\alpha)$
is defined in Ref. \cite{experiments} as
\begin{equation}
 \kappa^{e}(\alpha)=\eta_{0}\frac{v_{rms}}{dp_{rms}/dz}
\end{equation}

Hence, since the velocity in the experiment varies sinusoidally, the above expression assumes the form   
\begin{equation}
 \kappa^{e}(\alpha)=\eta_{0}\frac{v_0}{\rho z_0 w^{2}}
\label{ldrexp}
\end{equation}
where $v_0$ denotes the amplitude of the velocity oscillation.

Thus, the absolute value of (\ref{localdynamic}) can be compared directly
 with (\ref{ldrexp}), provided the velocity and the pressure gradient are sinusoidals.

  Theoretical results given by the Maxwell model (\cite{experiments},\cite{pre2},\cite{Maxwellteorico}),  reproduce quite well the values of $\alpha$ at which the peaks in the \textbf{LDR} of the \textit{CPyCl/NaSal} solution are observed. However, the  predicted relative amplitude deviates from the experimental results. In the Maxwell model the relative amplitude is larger than the experimental values and this model also fails in predicting the decrease of the \textbf{LDR} observed experimentally.  In Ref.\cite{experiments} it is speculated that these differences might be due to  nonlinear phenomena that occur in the \textit{CPyCl/NaSal} solution as well as to compressibility  effects that could take place when the frequency is increased. Regarding the nonlinearities, as it was mentioned in the \textit{Introduction}, there are two potential sources of them in the hydrodynamic equations, inertial and elastic. The former source is vanishingly small, since the Reynolds number is set $R_e\ll 10^{-4}$ (see \cite{experiments}). As for the elastic sources of nonlinearity we must also say that, in \cite{pre2}, measurements were carried out at a more detailed level, namely, the whole velocity field was explored along the cross-section and at different heights from the moving piston. The former measurements confirm the hypothesis (to a certain extent) that the velocity field depends only on the radial coordinate. With this assertion in mind, let us now examine a little more closely the Oldroyd-B constitutive equation (Eq. \ref{ecconstupper}) in the case of pure shear flows.

  Let us first note that in such flows the velocity gradient tensor satisfies, besides the incompressibility condition $Tr{D}=0$, an additional constraint $\left(\nabla v\right)_{ij}\left(\nabla v\right)_{jk}=0$ \cite{polymerstresspol}. Moreover, the operator $\vec{v}\cdot\nabla$ vanishes identically, as mentioned above. Then, it follows for the upper convected derivative of the velocity gradient tensor

    \begin{equation}
     \frac{\delta D}{\delta t}=\frac{\partial D}{\partial t}-2 (\nabla v)^t \cdot\nabla v.
      \label{uppershearflow}
    \end{equation}

 With the latter result, the Oldroyd-B constitutive equation can be written as 
    \begin{equation}
\sigma +\lambda\dfrac{\partial \sigma}{\partial t}-\lambda\left[\sigma\cdot\nabla v+ (\nabla v)^t\cdot \sigma\right]=
\eta_0\left(\mathbf{D}+\lambda\frac{\eta_{s}}{\eta_0}\left[\frac{\partial D}{\partial t}-2 (\nabla v)^t \cdot\nabla v\right]\right).
     \label{polymstressshear}
          \end{equation}

 The effects (assumed weaks) of the nonlinear terms can be assessed by a perturbation procedure. Indeed, an expansion of the stress tensor in terms of the Weissenberg number \footnote{In oscillating flows we can define the Weissenberg number as $Wi= \lambda\gamma_o$, where $\gamma_o$ is the root mean square of the characteristic velocity gradient} $\sigma=\sigma_0+ (Wi) \sigma_1+O(Wi^2)$ substituted in Eq. \ref{polymstressshear} gives to zero and first order 

      \begin{equation}
       \sigma_0+\lambda\dfrac{\partial \sigma_0}{\partial t}=2
\eta_0(\nabla v+(\nabla v)^t)
\label{pertordencero}
      \end{equation}
and
       \begin{equation} 
 \sigma_1+\lambda\dfrac{\partial \sigma_1}{\partial t}=\eta_s \dfrac{\partial (\nabla^* v^*+ (\nabla^* v^*)^t)}{\partial t}-2\eta_s \gamma_o\nabla^*v^*\cdot(\nabla^* v^*)^t+\left[\sigma_0\cdot\nabla^* v^*+ (\nabla^* v^*)^t\cdot \sigma_0\right],
\label{pertordenuno}
     \end{equation}
respectively. In equation \ref{pertordenuno}, the quantity  $\nabla^*v^*$ denotes a dimensionless velocity gradient tensor defined by the relation $\nabla v=\gamma_o\nabla^*v^*$. In the particular case we are concerned with, in which the only non-vanishing component of the velocity gradient is ${\nabla v}_{rz}=\frac{\partial v_z}{\partial r}$, from Eqs. \ref{pertordencero} and \ref{pertordenuno} we can see that the last two terms in the r.h.s of \ref{pertordenuno} are responsible for the appearance of a normal stress $\sigma_{zz}$. This shear-induced anisotropy in the stress tensor is a nonlinear effect which is absent from the corresponding Newtonian flow and has been disregarded in our present derivations.  Due to this anisotropy, if the fluid suffered a perturbation of large enough amplitude it would have a nonlinear (subcritical) transition to a sort of weakly turbulent state \cite{reports}. In such a case our analytical treatment will break down. However, from the analysis of the order of magnitudes of the terms involved in the equations resulting from our perturbation procedure, we can note that shear stresses becomes much larger than the normal ones when the following condition is satisfied $\frac{\gamma_o}{\omega}\ll 1$. Then, the replacement of the upper-convected time derivative by the local time derivative seems to be a good approximation at high frequencies and indeed, as we shall see, our linear theory provides a better fit at high frequencies.   Nevertheless, at low frequencies non linear effects must be taken into account and in our opinion, are the main cause of the evident disagreement with the predictions of both the linear Oldroyd-B and Maxwell models, as we shall show in the following. 

     In Figure 1 the comparison of our calculation with experimental results is shown.
 As in Ref. \cite{experiments} we plot the LDR
 scaled by its value at steady motion versus $\alpha$. The theoretical curve exhibits a 
 good agreement with experiments. The curve shape is 
quite similar for values of $\alpha$ in the range of 0.4 to 1 and only  slightly differs in
 the amplitude of the response. We can see that the decrease in the dimensionless LDR observed experimentally is quite well reproduced. This allows us to infer that as 
$\alpha$ increases, the purely Newtonian component becomes stronger and dissipative effects 
 are more evident. In fact, this can be predicted from equation (\ref{ecconstmod}). From the definition of $\alpha$ and 
assuming that the viscosity $\eta_o$ does not depend on the velocity gradients, an increase in $\alpha$ leads to
an increase in the number $\omega\lambda\frac{\eta_s}{\eta_0}$. If the latter number is much larger than unity
we have in equation (\ref{ecconstmod})
   \begin{equation}
    \|\lambda\frac{\partial \sigma}{\partial t}\|\gg\|\sigma\| \;\;\;\;\;\;\;
 \|\lambda\frac{\eta_s}{\eta_0}\frac{\partial D}{\partial t}\|\gg\|D\|.
   \end{equation}
Therefore, in this limit we have for the stress in the viscoelastic solution
   \begin{equation}
    \sigma\sim 2\eta_{s} D
   \end{equation}

Hence, the solution behaves in this case, as a Newtonian liquid. Moreover, from the above arguments we  see
that the Newtonian factor becomes stronger as the number  $\omega\lambda\frac{\eta_s}{\eta_0}$ 
is increased. This is a significant difference from Maxwell liquids, whose behavior is similar to a Hookean
solid when $\omega\lambda\gg1$ \cite{birdstewart}. Thus, there is a transition from viscoelastic to purely dissipative (Newtonian) dynamics of oscillating flows of Oldroyd-B fluids. We stress the fact that the relevant parameter for this transition is the number $\omega\lambda\frac{\eta_s}{\eta_0}$, which does not depend on the geometrical properties in which the fluid is confined.  Another noteworthy  observation is that, neither nonlinearities nor compressibility effects have been taken  into account.  Nevertheless, in Figure 1 it can be seen that both models exhibit similar behavior and overestimate the relative amplitude of the LDR at low frequencies. These disagreements, as mentioned above, might be caused by nonlinearities. 

  The curve for the Oldroyd-B model was obtained using the relaxation time, $\lambda=1.9s$, and the viscosity at 
steady flow, $\eta_0=60Pa\cdot s$ reported in Refs. (\cite{experiments},\cite{shearthinning}). Besides, the 
value of the solvent viscosity which provides the best fit to the experimental data was $\eta_s =0.08 Pa\cdot s$. 
 This value is approximately 80 times larger than the water viscosity, which is the solvent of the
\textit{CPyCl/NaSal} 60/100 solution.
 Thus, the Newtonian factor considered here within the Oldroyd-B model must be thought as a 
contribution leading to  an effective viscosity larger than the water viscosity. This result might be due to the fact that
  \textit{CPyCl/NaSal} 60/100 solution is not very dilute \cite{shearthinning}, whereas
 the Oldroyd-B model is valid only for dilute polymer solutions (\cite{larson}, \cite{berti}), where the hydrodynamic interactions are neglected. 
 Nevertheless, for highly concentrated solutions, disturbances of the solvent velocity
 field generated by the motion of the polymer molecule (coil) affects the drag force on the neighboring coils. As a consequence, the total drag force is larger than the total force  that would be
  produced if the solution were very dilute. So, using in this case the Oldroyd-B model leads to consider effective values of the parameters involved in the model.

\begin{figure}[h!t]
\begin{center}
 
\includegraphics[width=0.5\textwidth]{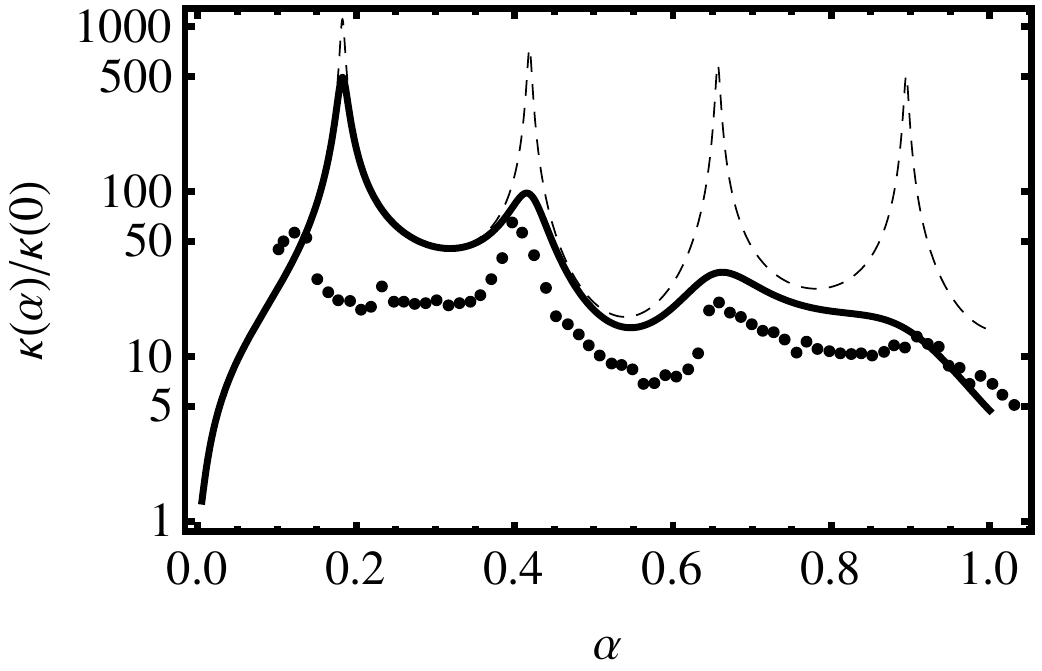}
\end{center}

\caption{Comparison between the Maxwell and Oldroyd-B models with experiments for the \textit{CPyCl/NaSal} solution .  
The dashed  line is the absolute
 value of $\kappa (\alpha)$ predicted by the Maxwell model. 
The continuous black line is the result for the Oldroyd-B
 model with the same value of A=173.71 and $\eta_0=60 Pa\cdot s$. The solvent viscosity used was $\eta_s=0.08 Pa\cdot s$, leading to  $\eta^{'}=0.8 \cdot 10^{-3}$.
Experimental values are shown by  points.}
\label{comparacion}
\end{figure}
Another quantity  reported in Ref. \cite{experiments} is the root mean square value of the velocity at the pipe axis..
Since, as mentioned before, the pressure gradient varies sinusoidally we assume that the velocity field has the form
\begin{equation}
 v_z(r,t)=Im(\phi(r)e^{i \omega t})
\label{rmssol}
\end{equation}

Substitution of equations (\ref{rmssol}) and (\ref{pressuremex}) in (\ref{linearnsdesp}) leads to
 an ordinary differential equation for $\phi(r)$\footnote{Actually, this is the same equation obtained for the Fourier transform of the velocity}, whose vanishing solution  at the pipe wall, $r=R$, is given by 
\begin{equation}
 \phi(r)=\frac{z_0 \omega}{i}\left(1-\frac{J_0(\beta r)}{J_0(\beta R)}\right),
\label{amplituder}
\end{equation}

 \begin{equation}
\beta^2=-i\frac{\omega}{\nu_0}\left(\frac{1+i\omega\lambda}{1+i\omega\lambda\eta_s/\eta_0 }\right).
\end{equation}

From equations (\ref{amplituder}) and (\ref{rmssol}) we have for the velocity
\begin{equation}
 v_z(r,t)=H_0(r,\omega)sin(\zeta+ \omega t)
\end{equation}
where $H_0(r,\omega)=\sqrt{Re^2(\phi(r))+Im^2(\phi(r))}$ and $\zeta=ArcTan\frac{Im(\phi(r))}{Re(\phi(r))}$
  
Therefore, the root mean square of the velocity can be calculated as 
 \begin{equation}
  V_{rms}=\frac{H_0(r,\omega)}{\sqrt{2}}
\label{definitionrms} 
 \end{equation}

The next figure shows a comparison between predictions of the Oldroyd-B model and experimental values for the root mean square velocity at $r=0$
\begin{figure}[h!t]
\begin{center}
 \includegraphics[width=0.5\textwidth]{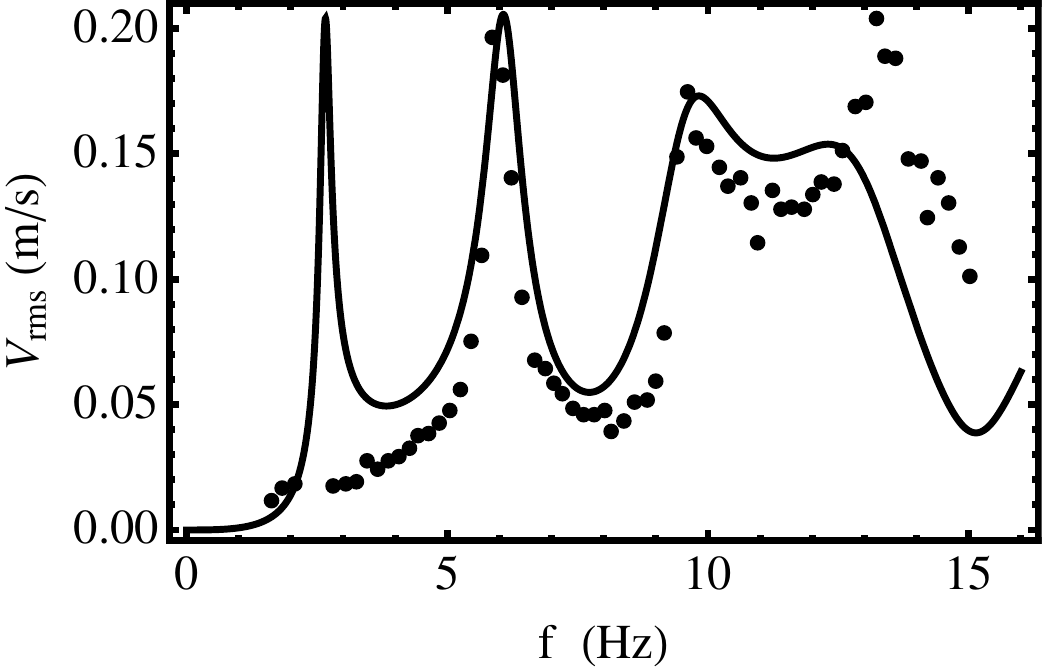}
\end{center}
\caption{Superposition of the root mean square velocity predicted by 
the Oldroyd-B model at the pipe axis represented by the continuous black line,
 equation (\ref{definitionrms}) evaluated at $r=0$, with the experimental values,
 shown by points in the graph. A  good fit is here exhibited for frequencies higher
 than $4$ Hz. 
 Peaks are quite well predicted and
only differences are observed in their magnitude.
 The values of density and viscosity of the \textit{CPyCl/NaSal} solution were used (\cite{experiments}, \cite{shearthinning}).
 The solvent viscosity that provides the fit is equal to $0.08$ $Pa\cdot s$   }
\label{rmscomparison}
\end{figure}

  Figure 2 shows that the Oldroyd-B model can reproduce accurately the frequencies at which
peaks of the root mean square velocity are observed. Moreover, the curve shape is well reproduced
 for frequencies above 4 Hz. These agreements reinforce the fact that the pure Newtonian contribution of the solvent
is very important in the dynamic behavior of the \textit{CPyCl/NaSal} solution. 

  As for the compressibility, from Figure 2 the highest velocity in the fluid $v_m$ is approximately equal to
$0.3 m/s$. Then, since the speed of sound in liquids is of order $v_s\sim 10^3 m/s$, we have for the Mach
number $M\sim 10^{-4}$. At such low Mach numbers, compressibility effects must be weak (\cite{Schlichting},\cite{landau}),  so the compressibility does not seem to be an important factor in the decay of the peaks. Let us present in the next section a detailed derivation that confirms this assertion.


 \section{Influence of Compressibility}

 As mentioned above, at these low Mach numbers compressible effects must be weak. Nonetheless, in this section we'll 
be aimed at modeling the problem of sound waves propagation in viscoelastic fluids flowing inside pipes. The influence of viscoelastic properties in sound wave propagation phenomena in infinite fluids and in wall bounded flows (inside straight pipes) has already been addressed in \cite{bjp} and \cite{rcf}, respectively.    

 Let us first revisit the stress modeling problem in compressible fluids. It is well known (see for instance, \cite{batchelor} and \cite{chorin}) that the motion in the neighborhood of any point consists of the superposition of:

 1- a uniform translation,

 2- a pure strain motion characterized by the rate of strain tensor $D=\nabla v+(\nabla v)^t$, which can be decomposed itself into an isotropic expansion (contraction) in which the rate of extension of all line elements is $\frac{1}{3}\nabla\cdot v$ and a straining motion without changes of volume characterized by $D_{ij}-\left(\frac{1}{3}\nabla\cdot v\right) \delta_{ij}$, and

 3- a rigid body rotation with local angular velocity given by $\frac{\nabla x u}{2}$

 We shall restrict ourselves, as in previous sections, to flows in which the Weissenberg number is sufficiently small so we shall not consider  shear-induced anisotropy in elastic stresses. Besides, we can assume a linear relationship between normal stresses and velocity derivatives by considering small oscillations of fluid particles in the wave motion. Moreover, as in the case of incompressible viscoelastic solutions, we shall consider that the stresses generated by the straining motion without changes in volume and by isotropic expansion relax on time scales $\lambda_1$ and $\lambda_2$, respectively. Then, the constitutive (linear) equations for the extra viscoelastic stresses may be written as follows

  \begin{equation}
\sigma_{1}^e+\lambda_{1}\dfrac{\partial\sigma_{1}^e }{\partial
t}=\eta_{1}\textbf{D}-\frac{2}{3}\eta_{1}\nabla\cdot \vec{v}\; \mathbf{I},
\label{tei}
\end{equation}
and
\begin{equation}
\sigma_{2}^e+\lambda_{2}\dfrac{\partial \sigma_{2}^e}{\partial
t}=\eta_{2}\nabla\cdot \vec{v}\; \mathbf{I},\label{tec}
\end{equation}
where $\mathbf{I}$ is the identity tensor of rank 2.
 The total viscoelastic stress $\sigma$ is obtained by summing $\sigma_{1}^e$ and $\sigma_{2}^e$.

  At this point, it is useful to note that for times larger than $\lambda_1$ the trace of the total viscoelastic stress satisfies the differential equation 

\begin{equation}
Tr \sigma+\lambda_{2}\dfrac{\partial Tr \sigma}{\partial
t}=3\eta_{2}\nabla\cdot\vec{v}\;\mathbf{I},\label{traza}
\end{equation}
 which coincides with the postulated equation for the trace of the stress tensor in Ref. \cite{bjp} when the displacements of fluid particles in  the wave are small compared with the wave length or equivalently, when the velocity of fluid particles in the wave is small compared with the velocity of sound.  

 The formal solution for the total viscoelastic stress can be written as 
 \begin{equation}
\sigma=\int_{0}^{\infty}ds[G(s)\mathrm{D}(t-s)+\{K(s)-1/3
G(s)\}\mathbf{I} \emph{Tr} \mathbf{D}(t-s)],
 \label{eccl}
\end{equation}
in which $$G(t)=(\eta_{1}/\lambda_{1})e^{-t/\lambda_{1}}$$ and

$$K(t)=(\eta_{2}/\lambda_{2})e^{-t/\lambda_{2}}$$ denote the relaxation functions of shear and normal stresses, respectively.

 Substituting this formal solution into the momentum balance equation leads to 
 \begin{equation}
\rho\left((\vec{v}\cdot\nabla)\vec{v}+\dfrac{\partial \vec{v}}{\partial t}\right)=-\nabla
p+\eta_{s}\nabla^{2}\vec{v}
+\int_{0}^{\infty}ds[G(s)\nabla^{2}\vec{v}+A(s)\nabla(\nabla\cdot
\vec{v})],
\label{ecogm}
\end{equation}
where $A(s)=K(s)+\frac{1}{3}G(s)$.

 Since we are considering that the velocity of fluid particles is small compared with the velocity of sound we can neglect the term $(\vec{v}\cdot\nabla)\vec{v}$ in the above equation. For the same reason, the relative changes in the fluid pressure and density should be small. Hence, we can write the latter magnitudes as

 $$p=p_o+p'\;\;\;\;\ \rho=\rho_o+\rho',$$
where $p_o$ and $\rho$ are the values of the pressure and density, respectively, in the unperturbed state, while the primed quantities refer to the oscillating part.  

 The equation of continuity in the linear approximation can be written as 
  \begin{equation}
\dfrac{\partial \rho'}{\partial t}+\rho_{0}\nabla \cdot\vec{v}=0.
\label{eccontinuidad}
\end{equation}

We may also note, since we are considering small deviations from the state of equilibrium, that the changes in the entropy $s$ are of second order of smallness \cite{landau}. In this approximation, to the first order of accuracy, we have that changes in the fluid pressure are only caused by variations in the fluid density. Then we can write $p\approx p(\rho)\label{pd}$. Hence the small variations of $p'$ are related to changes in $\rho'$ in the following way
  \begin{equation}
   p'=\left(\frac{\partial p'}{\partial \rho}\right)_s \rho'=c^2\rho',
  \end{equation}
where $c$ is the speed of sound.

 After substituting for $\rho'$ according to the above equation in \ref{eccontinuidad} we obtain
  \begin{equation}
\dfrac{\partial p'}{\partial t}+c^2 \rho_{0}\nabla \cdot\vec{v}=0.
\label{twounknown}
  \end{equation}

 With the equations \ref{twounknown} and \ref{ecogm} for the four unknowns $p'$ and $\vec{v}$ we are able to completely describe the sound wave. Moreover, in order to express the two unknowns in term of one of them it is convenient to take divergence in both sides of equation \ref{ecogm} and take the time derivative of equation \ref{twounknown}. Then, it can readily be obtained the following equation describing the small oscillations of pressure (it was also derived in \cite{bjp}) 
\begin{equation}
\rho_{0}\dfrac{\partial^{2} p}{\partial
t^{2}}=K_{x}\nabla^{2}p+\eta_{s}\dfrac{\partial }{\partial
t}\nabla^{2}p+\int_{0}^{\infty}ds M(s)\dfrac{\partial }{\partial
t}\nabla^{2}p(t-s), 
\label{presiones}
\end{equation}
 in which $K_x=\rho_o c^2$ is the compressibility modulus and $M(s)=K(s)+\frac{4}{3}G(s)$. Here and henceforward we shall omit the prime symbol to the small variations of pressure.

 With the same assumptions as in section $3.1$ concerning the flow geometry we shall consider that the velocity field in the sound wave can be described in terms of its component along the z axis, i. e. 
  
\begin{equation}
     \vec{v}=v_z(r,z,t)\mathbf{e_{z}}.
    \end{equation}

 As a consequence of the latter, the pressure will vary only along the pipe axis and will be constant along a the cross section.
\begin{equation}
 p=p(z).
\end{equation}

 A noteworthy observation is that the velocity must depend on $z$, owing to the compressibility, which represent the main difference with the flow of an incompressible liquid.
 
 Let us seek a solution of equations \ref{twounknown} and \ref{ecogm} in the form of a plane wave,

  \begin{equation}
 v_z(r,z,t)=\textit{Re}[\phi(r)e^{i (kz-wt)}]\;\;\;\;\
 p=\textit{Re}[P_{0}e^{i (kz-wt)}]. 
\label{velpresarmonicos}
 \end{equation}

 Substitution of the above expression for the pressure in equation \ref{twounknown} leads to the dispersion relation

  \begin{equation}
 k^{2}=\left(\frac{w}{c_{x}}\right)^{2}\left[1+\frac{i w \eta_{0}}
{\rho_{0}c_{x}^{2}}\left(\frac{\eta_{s}}{\eta_{0}}+\frac{4}{3}
\frac{\eta_{1}}{\eta_{0}}\frac{1}{1-i w \lambda}+\frac{\eta_{2}}
{\eta_{0}}\frac{1}{1-i w \lambda}\right)\right]^{-1}.
\label{ecuaciondedispersion}
\end{equation}

 Similarly, after substitution of the expression for the velocity and the pressure itself in \ref{ecogm} we obtain an ordinary differential equation for $\phi(r)$
 \begin{equation}
 \nabla^{2}_{r} \phi(r)+\left(\frac{i \omega \rho_{0}}{\eta_{0}}
\frac{1-i\omega\lambda}{1-i\omega\lambda\eta^{'}}-\frac{k^{2}
\left(\frac{4}{3}(1-\eta^{'})+\eta_{2}^{'}\right)}
{i\omega\lambda\eta^{'}}\right)\phi(r)=-i \frac{k P_{0}}{\eta_{0}}\frac{1-i\omega\lambda}{1-i\omega\lambda\eta^{'}},
\label{ecuacionphi(r)}
\end{equation}
 in which $\nabla^{2}_{r}=\dfrac{1}{r}\dfrac{\partial}{\partial r}
\left(r \dfrac{\partial }{\partial r}\right)$, $\eta^{'}=\frac{\eta_s}{\eta_0}$ and $\eta_2^{'}=\frac{\eta_2}{\eta_0}$.

 Solving the equation for $\phi(r)$ we obtain for the velocity
\begin{equation}
 v_{z}(r,z,t)=-\frac{i k P_{0}e^{i(kz-wt)}}{i\omega\rho_{o}-\frac{k^{2}\eta_{0}}{1-i\omega\lambda}\left(\frac{4}{3}(1-\eta^{'})+\eta_{2}^{'}\right)}
\left(1-\frac{J_{0}(\beta_{c}r)}{J_{0}(\beta_{c}R)}\right).
\label{velocidaddependencias}
\end{equation}
In the above expression it has been defined $$\beta_c^{2}=\left(\frac{i \omega \rho_{0}}{\eta_{0}}
\frac{1-i\omega\lambda}{1-i\omega\lambda\eta^{'}}-\frac{k^{2}
\left(\frac{4}{3}(1-\eta^{'})+\eta_{2}^{'}\right)}
{i\omega\lambda\eta^{'}}\right).$$

 Noting that $\frac{d p}{d z}=i k P_{0}e^{i(kz-wt)}$, we can conveniently rewrite the above equation as

 \begin{equation}
 v_{z}(r,z,t)=-\frac{1}{i\omega\rho_{o}-\frac{k^{2}\eta_{0}}{1-i\omega\lambda}\left(\frac{4}{3}(1-\eta^{'})+\eta_{2}^{'}\right)}
\left(1-\frac{J_{0}(\beta_{c}r)}{J_{0}(\beta_{c}R)}\right)\frac{\partial p}{\partial z},
\label{velocidadgradpresiones}
\end{equation}
which resembles the corresponding velocity for the incompressible case (Eq. \ref{vincompresible}). 

 Hence, defining the local dynamic permeability as in section $3.2$ by
\begin{equation}
 \kappa_c=\frac{-\eta_0 v_{z}(0,z,t) }{\frac{\partial p}{\partial z}},
\label{permeabilidadlocal}
\end{equation}
 and referring the frequency to $\frac{\nu_0}{R^2}$, the inverse of the characteristic time of vorticity diffusion due to viscosity, it is obtained  

\begin{equation}
 \kappa_c(\alpha)=-\frac{R^2}{i \alpha-\varepsilon_{c}}\left(1-\frac{1}{J_{0}(\beta_{c}^{'})}\right),
\label{locapermcompresible}
\end{equation}
in which 
\begin{equation}
 \varepsilon_{c}=\frac{\left(\frac{\alpha}{Re_{x}}\right)^{2}\left[\frac{4}{3}(1-\eta^{'})+\eta_{2}^{'}\right]}{\left(1-i\alpha A\right)\left(1+i\frac{\alpha}{Re_{x}^2}\left(\left[\frac{4}{3}(1-\eta^{'})+\eta_{2}^{'}\right]\frac{1}{1-i \alpha A}+\eta^{'}\right)\right)}
\label{relacionesdispersionadime} 
\end{equation}
and
\begin{equation}
 \beta_c^{'2}=\left(i\alpha -\varepsilon_c\right)\frac{1-i\alpha A}{1-i\alpha A\eta^{'}}
\label{argumentocompresible}
\end{equation}

In the above expressions it has been used $Re_x=\frac{ c_x R}{\nu_o}$. The dimensionless numbers $A$ and $\alpha$ are defined in the same way as before. 

In experiments with the  $CPyCl/NaSal$ solution,
 the range of frequencies covered kept the number $\alpha$ in the range
 $0<\alpha<1$. Estimating that the speed of sound in the
 solution is $c_x\sim 10^{3}$ $m/s$, leads to $Re_x\sim 10^{3}$ and thus
 $\left(\frac{\alpha}{Re_{x}}\right)^{2}\sim 10^{-6}$.
 Let us also assume that $\eta_2'\sim 1$ (see \cite{landau}) and that relaxation times of shear and normal stresses are of the same order of magnitude. Evaluating the remainder parameters as in the incompressible case we obtain that $Abs(\varepsilon_c)\ll1$ leading to $\beta_{c}^{'2}\cong\beta^{'2}$. Thus, we reach the conclusion that for weakly compressible flows the $\mathbf{LDR}$ matches almost perfectly the corresponding incompressible flow. We can see very slight differences in the values of frequencies at which peaks are observed as well as in the amplitude of the $\mathbf{LDR}$ provided only that $\frac{\alpha}{Re_x}\sim 1$ \cite{rcf}.

\section{Conclusions}
In this paper we have discussed some non-Newtonian effects of an Oldroyd-B fluid
  under oscillating pressure gradient. A relation was derived  between flux velocity
  and pressure gradient in frequency domain that resembles Darcy's Law.
 An analytical expression was given for the dynamic permeability in terms of 
 dimensionless parameters, which are: the number $\alpha$ that represents the 
 ratio between the inertia and viscosity forces, the inverse of Deborah's number,
  which is the ratio between characteristic times of viscous effects and elastic
  ones, and the number $\omega\lambda\frac{\eta_s}{\eta_0}$,
  which determines the differences between the Oldroyd-B and Maxwell models. Moreover, our calculations have predicted a transition from a Non-Newtonian (viscoelastic) to purely Newtonian behavior in the dynamics of viscoelastic solutions when the condition $\omega\lambda\frac{\eta_s}{\eta_0}\gg 1$ is satisfied. This transition is therefore independent of the flow geometry, i.e. it has a universal character.  We have found good agreement between the linear Oldroyd-B model predictions for the local dynamic response, as well as for the root mean square velocity at the pipe center,  and the corresponding experimental data, shown in Figure 1 and 2,  respectively. This suggests that it is the dissipative factor, present in Oldroyd-B fluids, which might leads to the decrease of the local dynamic response observed  in experiments \cite{experiments}. However, these good agreements are obtained when the solvent viscosity used in the fitting is considered as an increased coefficient  owing to the possible fact that due to high concentrations of the viscoelastic solution, disturbances of the solvent velocity
 field generated by the motion of the polymer molecule (coil) affects the drag force on the neighboring coils. Consequently, the solvent viscosity is used here as a free parameter actually. The very small Mach numbers at which experiments were made indicate that compressibility effects can be neglected. Our derivations confirmed that, given the experimental conditions, weakly compressible effects won't contribute to any significantly change in the local dynamic response of the fluid compared against the incompressible result. However, discrepancies are evident at low frequencies in both figures.  These disagreements in this range of frequencies could be
 produced by nonlinearities as the perturbation procedure developed suggested.  According to our results, anisotropy in viscoelastic stresses should be stronger at low frequencies. The latter would drive the viscoelastic pipe flow weakly turbulent (through a nonlinear subcritical transition) if the fluid were perturbed significantly and consequently, in such state our hypotheses will break down.       
   
------------------------------------------------------------

\bibliographystyle{amsplain}

\begin{thebibliography}{99}
\bibitem{Schlichting} Hermann Schlichting \textit{Boundary Layer Theory} Mc Graw-Hill Book Company 1973.
\bibitem{rcf} Anier Hern\'andez Garc\'ia and Oscar-Sotolongo Costa. \textit{Cuban Journal of Physics}, Vol. 28, No. 1, (2011).\;\;\;\;\ Anier Hern\'andez Garc\'ia.\;\;\ \textit{ Diploma thesis}   "Waves in Viscoelastic Media", (2009).
\bibitem{enhacement}  D. Tsiklauri and I. Beresnev. \textit{Phys. Rev. E} 63,046304 (2001)
\bibitem{experiments} J. R. Castrej\'on-Pita, J. A. del R\'io, A. A. Castrej\'on-Pita, and G.
Huelsz.\;\;\;\;\ \textit{Phys. Rev. E} 68, 046301 (2003)
\bibitem{pre2} M.Torralba, J. R. Castrej\'on-Pita, A. A. Castrej\'on-Pita, G.
Huelsz, J. A. del R\'io,   \textit{Phys. Rev. E} 72, 016308 (2005)
\bibitem{pre1}  J. A. del R\'io, M. L\'opez de Haro, and S. Whitaker\;\;\;\;\ \textit{Phys. Rev. E} 58, 6323 (1998)
\bibitem{Oldroyd} J.G. Oldroyd,\textit{ On the formulation of rheological equations of state},\;\;\;\ Proc. R. Soc.
A 200 (1950) 523–541
\bibitem{Skelland} Skelland \;\;\;\;\ \textit{Non Newtonian Flow and Heat Transfer} Wiley, New York, 1967 
\bibitem{birdstewart} R.B. Bird, W.E. Stewart and E.N. Lightfoot, \;\;\;\;\ \textit{Transport Phenomena}, (John Wiley  Sons, Second Edition, 2002).

\bibitem{larsonins} S.J. Muller, R.G. Larson and E. S. G. Shaqfeh
, \textit{Rheol. Acta} 28: 499-503 (1989)

\bibitem{Prilutski} Prilutski G., Gupta R. K., Sridhar T. and Ryan M. E.   \textit{J. Non-Newton. Fluid Mech.} \textbf{12},233-41, 1983.

\bibitem{larson} R.G. Larson \;\;\;\;\ \textit{The Structure and Rheology of Complex Fluids},  (Oxford University Press, 1999).
\bibitem{Maxwellteorico} J.A. del R\'io, J.R. Castrej\'on-Pita \textit{Revista Mexicana de F\'isica} 49 (1) 74-85.

\bibitem{polymerstresspol}  Victor S. L’vov, Anna Pomyalov, Itamar Procaccia, and Vasil Tiberkevich  \textit{Phys. Rev. E} 71, 016305 (2005) 
\bibitem{reports}  Alexander N. Morozov, Wim van Saarlos  \textit{Physics Reports} 447 (2007), 112-143
\bibitem{shearthinning} J.F.Berret, J. Apell, and G. Porte, Langmuir 9, 2851,(1993)

\bibitem{berti} S. Berti, \textbf{Ph.D. Thesis} \textit{Non-Newtonian turbulence: viscoelastic fluids and binary mixtures. }, 2006
\bibitem{bjp} Oscar Sotolongo Costa, Alexei V\'azquez V\'azquez, and Jos\'e Mar\'in Antu\~na.\;\;\;\;\ 
\textit{Brazilian Journal of Physics}, vol. 27, no. 3
\bibitem{landau} L.D.Landau and E.M.Lifshitz.\;\;\;\;\ \textit{Fluid
Mechanics} (1987 Pergamon Books Ltd.)
\bibitem{batchelor} G. K. Batchelor, \;\;\;\;\ \textit{An introduction to fluid dynamics} (Cambridge University Press, 2000)
\bibitem{chorin} A. J. Chorin and J. E. Marsden\;\;\;\;\ \textit{A mathematical introduction to fluid mechanics} (1990, 1993 Springer-Verlag, New York Inc)

\end{thebibliography}

\end{document}